
\magnification=1200
\hsize=15truecm
\vsize=23 truecm
\baselineskip 20 truept
\parindent=1cm
\overfullrule=0pt
\font\abs=cmr10
\def\pmb#1{\leavevmode\setbox0=\hbox{$#1$}\kern-.025em\copy0\kern-\wd0
\kern-.05em\copy0\kern-\wd0\kern-.025em\raise.0433em\box0}
\noindent
CERN-TH.6777/93
\noindent
POLFIS-TH.24/93
\vskip 2truecm
\centerline
{\bf STRING QUANTUM SYMMETRIES FROM PICARD FUCHS}
\centerline
{\bf EQUATIONS AND THEIR MONODROMY $^*$}
\vskip 1truecm
\centerline{R. D'AURIA}
\centerline{\it Department of Physics, Politecnico di Torino}
\centerline{\it and INFN Sezione di Torino, Torino, Italy}
\vskip 10pt
\centerline{S. FERRARA}
\centerline{\it CERN, 1211 Geneve 23, Switzerland}
\vskip 1truecm
\midinsert
\centerline{ABSTRACT $^{**}$}
\vskip 1truecm
\baselineskip 14pt
\narrower{\noindent\abs
Local and global properties of the moduli space of
Calabi--Yau type compactifications determine the low energy parameters
of the string effective action. We show that the moduli space geometry
is entirely encoded in the Picard--Fuchs equations for the periods
of the Calabi--Yau $H^{(3)}$--cohomology.}
\endinsert
\vskip 1truecm
\noindent
$^*$ Supported in part by DOE grants DE-AC0381-ER500050,\qquad
DOE-AT03-88ER40384, task E and by MURST.

\vskip 1truecm

\vfill\eject

\noindent
{\bf 1.\ Introduction}
\vskip 10pt
\noindent
It is well known that among the possible vacua of heterotic string
theory the  (2,2), c=9 world--sheet superconformal field theories
(SCFT) are of prominent importance since they give rise to N=1
supersymmetry in four dimensions, which is currently believed to be
a necessary ingredient of the low--energy effective Lagrangian
$^1$.

In general it is quite difficult to compute the correlators of a
Superconformal Field Theory
(SCFT) and usually more insight into their structure can be gained
by using the Landau--Ginzburg approach to N=2 SCFT $^2$ or else in
the context of the N=2 topological field theory $^3$. The description
of the (2,2)--string vacua in terms of the Landau--Ginzburg
superpotential (possibly in its twisted topological version) is
strictly related to the geometrical description in terms of the
compactification of the heterotic string on a Calaby--Yau 3--fold
(C.Y.)
and in fact the determination of the physical low--energy parameters in
both frameworks can be performed by using the same techniques of
algebraic geometry $^{4,5}$. One is actually faced with the problem of
computing
such low--energy quantities in terms of the properties of the moduli
space of a given SCFT. Moduli space is the space of all the marginal
deformations of the underlying N=2 SCFT or, in the geometrical picture of
Calabi--Yau compactifications, is the space of the parameters
$\varphi^\alpha$, $\psi^a$ describing respectively the deformations of
the K\"ahler class and/or of the complex structure of the C.Y. manifold
$^6$.
In the low energy theory the moduli parameters $\varphi^\alpha(x)$,
$\psi^a(x)$ appear as neutral massless scalar fields with vanishing
potential.

The low energy objects that we are interested in as functions of the
moduli are: i) the moduli and family metrics; ii) the Yukawa couplings.
Another important moduli dependent effect is: iii) the running of the
gauge couplings due to one--loop stringy effects.

These quantities appear in the effective N=1 Lagrangian in the
following form:

\noindent
i)

$$
g_{i\bar j} \partial_\mu  M^i \partial_\nu \bar M^{\bar j}
G^{\mu\nu}\eqno(1.1)
$$

$$
g^{(27)}_{\alpha\bar\beta} \partial_\mu \phi^\alpha_{27}
\partial_\nu \bar\phi^{\bar\beta}_{27} G^{\mu\nu}\eqno(1.2)
$$

$$
g^{(\overline{27})}_{a\bar b} \partial_\mu \phi^a_{\overline{27}} \partial_\nu
\bar\phi^{\bar b}_{\overline{27}} G^{\mu\nu}\eqno(1.3)
$$

\smallskip

\noindent
ii)

$$
W_{\alpha\beta\gamma}(\varphi) \phi^\alpha_{27} \phi^\beta_{27}
\phi^\gamma_{27}
\eqno(1.4)
$$

$$
W_{abc} (\psi) \phi^a_{\overline{27}}
\phi^b_{\overline{27}}\phi^c_{\overline{27}}\eqno(1.5)
$$

\smallskip
\noindent
iii)

$$
{1\over g^2(S,\varphi,\psi)} T_r F_{\mu\nu} F^{\mu\nu} +
\theta(S,\varphi,\psi) T_r F_{\mu\nu} \tilde F^{\mu\nu}
\eqno(1.6)
$$

\noindent
where $M^i$ denote any of the two sets of moduli $\varphi^\alpha$,
$\psi^a$, $S=D+ia$ is the dilaton--axion scalar field,
$g_{i\bar j}, g^{(27)}_{\alpha\bar\beta}, g^{(27)}_{a\bar b}$
are the metrics for moduli, families and antifamilies respectively,
$W_{abc}$, $W_{\alpha\beta\gamma}$ are the Yukawa couplings, $g^2$
and $\theta$ are the gauge coupling and the $\theta$--angle for the
gauge field strength $F_{\mu\nu}$ ($\tilde F_{\mu\nu}$ is the dual of
$F_{\mu\nu}$).

We note that the moduli space ${\cal M}$ has the structure of a
product space

$$
{\cal M} = {\cal M}_1\times {\cal M}_2\eqno(1.7)
$$

\noindent
where ${\cal M}_1, {\cal M}_2$ are parametrized by the $\varphi^\alpha$
and $\psi^a$, respectively. This can be proven either in SCFT or
by supersymmetry arguments $^7$. The two spaces are actually related by the
mirror hypothesis $^8$ which establishes that exchanging the role of the
two parameters $\varphi^\alpha$, $\psi^a$ associated to the K\"ahler
class and complex structure deformations of the C.Y. 3--fold,one obtains
a mirror C.Y.-$3$-fold in which the $H^{(2)}$ and $H^{(3)}$ cohomology
classes are exchanged,due to the isomorphism of the underlying SCFT's
under sign change of the $U(1)$-charge.
 In the following we shall restrict our attention only to
the manifold ${\cal M}$ of the complex structure moduli. The results
thus obtained are also valid for the moduli space of the K\"ahler class
deformations of the associated mirror C.Y. manifold.

The structure of the quantities appearing in eq.s (1.1--6) is
completely determined by the local and global properties of the
moduli space. These properties in turn are encoded in the
Picard--Fuchs equations for the periods $X^A(\psi), F_A(\psi)$
of the Calabi--Yau family parametrized by the $\psi^{a'}$s.
Once we know the periods we may reconstruct the K\"ahler
potential of the moduli space and the Yukawa couplings as follows
$^6$:

$$
K(\psi,\bar\psi) = - \log\ i(X^A \bar F_A - \bar X^A F_A)
\eqno(1.8)
$$

$$
W_{abc} = {\partial X^A\over\partial\psi^a}
{\partial X^B\over\partial\psi^b}
{\partial X^C\over\partial\psi^c}
{\partial^3 F\over\partial X^A \partial X^B \partial X^C}
\eqno(1.9)
$$

\noindent
where $F=F(X^A)$ is a degree two homogeneous function such that

$$
X^A F_A = 2F\quad\rightarrow\quad
F_A \equiv {\partial F\over\partial X^A}\eqno(1.10)
$$

\noindent
Notice that the characterization of the local geometry of the
moduli space given by eq.s (1.8--1.10) coincides with the
definition of Special K\"ahler $^{9,10,11}$ geometry and indeed the
periods $(X^A, F_A)$ of the Calabi--Yau 3--fold can be indentified
with a set of globally defined holomorphic sections in terms of
which the special K\"ahler geometry can be defined. In this
lecture however we will be mainly concerned with the global
properties of the moduli space ${\cal M}$
which are consequence of the fact that
${\cal M}$ possesses a group $\Gamma$ of discrete isometries which
is generally referred to as the target space duality group, or
modular group $^{12}$. The duality group describes quantum symmetries
of the string effective actions and is the discrete version of
the non--compact symmetries of old supergravity Lagrangians
(no--scale supergravities). The most celebrated example is
the effective Lagrangian obtained by compactification of the
heterotic string on a 6--torus, modded out by some discrete symmetry ( orbifold
) with residual $N=1$ space-time supersymmetry. The K\"ahler class untwisted
modulus (corresponding to the volume size)
$t=2(R^2+i\sqrt{|g|})$
parametrizes the homogeneous space $^{13}$:

$$
{SU(1,1)\over U(1)}\eqno(1.11)
$$

This sector of the theory is obviously invariant under $PSL(2,\bf R)$
The K\"ahler potential of the scalar fields $t$,$\phi_{27}$ is

$$
K=-3\log\ [i (t-\bar t)-\phi_{27} \bar\phi_{\overline{27}}]\eqno(1.12)
$$

\noindent
and the superpotential ${\cal W}$ is given by ${\cal W}=C\phi^3_{27}$
corresponding to a constant Yukawa coupling, $W=C$. The theory
is obviously invariant under

$$
t' = {at+b\over ct+d}\quad , \quad
\phi'_{27} = - {1\over ct+d} \phi_{27}\quad , \quad
ad-bc=1\eqno(1.13)
$$

\noindent
However,when other (twisted) sectors are introduced and/or quantum corrections
computed, the duality (quantum symmetry) group is given by $SL(2,{\bf Z})$
with generators

$$
t  \rightarrow t+1\qquad\qquad
t  \rightarrow {-1\over t}\eqno(1.14)
$$

\noindent
For Calabi-Yau manifolds $W$ is no longer a constant, but in
general it depends on the moduli : $W_{\alpha\beta\gamma} =
W_{\alpha\beta\gamma} (M)$ because of the instanton corrections
to the $\sigma$--model perturbative result ( large radius limit). In the one
modulus
case, however, it turns out that $t\rightarrow t+1$ is still an
exact symmetry if $t$ is the special coordinate of the
K\"ahler Special Geometry, or, equivalently, the flat coordinate
of the associated topological field theory. Actually one is led
to consider the possibility that the translational symmetry is
exact for any number of moduli since invariance under
$t^a\rightarrow t^a+1$, $a=1,...,n$  has its stringy origin in
the presence of the antisymmetric axion field $B_{ij}$ in the
$\sigma$--model action, and means that the $t^a$--variables
are periodic. For example the Yukawa coupling in the one modulus case
can be written as follows

$$
W(t) = \sum^\infty_{n=0} d_n\ e^{2\pi int}\eqno(1.15)
$$

\noindent
and for many variables we have analogous expansions. In the
large radius limit, $t\rightarrow i \infty$, eq. (1.15) gives

$$
W(t) = d_0 = const.\eqno(1.16)
$$

\noindent
which implies that ${\cal F}(t)\equiv |X^0|^{-2} F(X^A)$ is
cubic in $t$. Actually the most general prepotential such
that $W(t)$ and the K\"ahler potential (1.8) are invariant under
$t\rightarrow t+1$ is (Cecotti et al.$^7$)

$$
F(X^A) = (X^0)^2 [t^3 + \lambda(t)+ P_2(t)]\eqno(1.17)
$$

\noindent
where $P_2(t)$ is a polynomial of degree two with purely real
coefficients and $\lambda(t)$ a periodic function of $t$. If instead
invariance under arbitrary shifts is required, $t\rightarrow t+c$,
$c\in{\bf R}$, then $\lambda$ is a constant.
\noindent
Note that a non vanishing $\lambda$ is generated in $\sigma$-model perturbation
theory at the four loop level$^{4}$.
Under the inversion $t\rightarrow\displaystyle{{-1\over t}}$ the
K\"ahler potential transforms as follows

$$
K\rightarrow K + f(t) +f (\bar t)\eqno(1.18)
$$

\noindent
so that $W$, which has a non trivial K\"ahler weight undergoes
a non trivial transformation.

In the next section we will show that the generalization of the
inversion  for a generic duality group is given by

$$
t^a \rightarrow f^a (t^a,\partial_a{\cal F}, {\cal F}; A,B,C,D)\qquad
a=1,...,n\eqno(1.19)
$$

\noindent
where $n$ is the number of moduli, ${\cal F}{(t^a)}=(X^0)^{-2}F(X^A)$
 and $A,B,C,D$ are $(n+1)\times
(n+1)$ matrices parametrizing a generic $Sp(2n+2;{\bf Z})$
transformation  $M$:

$$
M=\left(\matrix{ A & B\cr C & D\cr}\right)\quad
\in\quad Sp(2n+2;{\bf Z})\eqno(1.20)
$$

For example in the case of the quintic studied by Candelas et al.
the duality group acting on the single modulus $t$ is generated by
two transformations:

$$
t\rightarrow t+1\eqno(1.21)
$$

$$
t\rightarrow {t\over t{\cal F}'_t - 2{\cal F}+1}\eqno(1.22)
$$

The knowledge of the duality group is very important since the
physical quantities appearing in the effective Lagrangian must
transform in a definite way under the group: for example the
gauge coupling $g^{-2}_a(t,\bar t)$ is a real function which
must be modular invariant in $t,\bar t$.

Moreover some of the duality transformations may induce
target--space anomalies in the effective Lagrangian $^{14}$.
These anomalies arise when the K\"ahler potential is not invariant
under the duality transformations, but undergoes the transformation
(1.18). In this case all the K\"ahler gauge dependent quantities
get transformed and anomalies can be generated. From the above
discussion we see that in general possibly anomalous transformations
belong to the coset ${\Gamma\over {\cal T}}$ where $\Gamma$ is the
duality group and ${\cal T}$ generates the discrete translations under which
the K\"ahler potential is invariant.
The modular group described here acts on the special coordinates
$t^a$ which are the coordinates which flatten the holomorphic
connection of the associated topological field theory. The
modular coordinate $\gamma$ on which the two transformations described
earlier act as $SL(2,R)$ transformations (for any $F$--function) is
described in the next section and the relation of this coordinate
with the special coordinate as well as with the Landau--Ginzburg
coordinate of the defining polynomial in CP$(d+1)$ will also be
given. In this respect the target space duality group $\Gamma$ is
closely related to the monodromy group $\Gamma_M$ of the Picard--Fuchs
equations which determine the period matrix for Calabi--Yau manifolds,
the precise relation  being discussed in the following.

\vskip 20pt
\noindent
{\bf 2.\ Monodromy of the Picard-Fuchs equations and the quantum
modular group}
\vskip 10pt
\noindent
The periods $(X^A, F_A)$ of a given family of Calabi--Yau d--folds
are known to satisfy a coupled set of linear partial differential
equations called Picard--Fuchs equations (PFE). A general property
of these equations is that they are of Fuchsian type, that is
they only have regular singular points. They can be derived from the
defining polynomial equation $W(y^i, \psi^\alpha)=0$ of the
Calabi--Yau manifold by using simple algorithms which have been
described in ref. [15]. If one solves the PFE for the periods
and uses eq.s (1.8) and (1.9), then one reconstructs the K\"ahler
potential of the moduli space and the Yukawa couplings associated
to that particular class of C.Y. compactifications. Furthermore
the PFE also encode essential informations on the global structure
of the moduli space through their monodromy group.

Let us denote by $\Gamma$ the target space duality group (quantum modular
group) and by $\Gamma_W$ the group of invariance of the superpotential
$W(y^i,\psi^a)$. $\Gamma_W$ consists of these diffeomorphisms of
the moduli $\psi^\alpha$ which leave $W=0$ invariant except for a
(quasi)--homogeneous change of the $CP(d+1)$ coordinates:

$$
W(y^i,\psi^\alpha)=0\qquad
{\mathop{\longrightarrow}\limits_{\Gamma_W}}\qquad
W(\tilde y^i(y);\tilde\psi^\alpha(\psi))=0\eqno(2.1)
$$

\noindent
where $\tilde y^i=U^i_ j y^j$ and $i,j$ run over all chiral fields
with same $U(1)$ charge.

Finally let us denote by $\Gamma_M$ the monodromy group of the
PFE's. To define it in the simplest way we restrict our
attention to the case of one single modulus, in which case the
PFE's are ordinary differential equations.

Then, if we denote by $(f_1(z),...,f_n(z))$ a basis of solutions
of the differential equation at a point $z$, by analytically
continuing $(f_1,...,f_n)$ along a closed loop around a
singularity $z_1$ of the equation we arrive at a new solution at $z_1$
which must therefore be expressible as a linear combination of the
basis $(f_1,...,f_n)$:

$$
(f_1,...,f_n)\ \rightarrow\
(\hat f_1,...,\hat f_n) =
(f_1,...,f_n)\ A_{z_1}\eqno(2.2)
$$

\noindent
where the $n\times n$ non singular matrix $A_{z_1}$ defines the
monodromy around $z_1$. If the equation has $r$ singular points we
obtain $r$ monodromy matrices $A_{z_1},...,A_{z_r}$, and if we
compose closed loops around $z_i$ and $z_j$ in the usual way it
is clearly seen that to the loop $\gamma_i\circ\gamma_j\equiv
\gamma_{ij}$ encircling $z_i$ and $z_j$ corresponds the monodromy
matrix $A_{z_j}\cdot A_{z_i}$, and that more generally
$A_{z_1},...,A_{z_r}$ generate a group, the monodromy group of the
differential equation (here the inverse $A^{-1}_{z_i}$ is the matrix
obtained by running around $z_i$ in the opposite direction, and {\bf 1}
corresponds to a circuit contractible to a point).

It turns out that in the known cases the monodromy group $\Gamma_M$
is a normal (generally infinite) subgroup of $\Gamma$ and that

$$
\Gamma/\Gamma_M \simeq \Gamma_W\eqno(2.3)
$$

Following the proposal of Lerche et al.$^{15}$ we assume that (2.3)
is true in general, possibly also for the case of many moduli.
Equation (2.3) suggests that in order to reconstruct $\Gamma$ we
can compute the monodromy group of the PFE's and the invariance
group of $W=0$, so that $\Gamma\simeq\Gamma_W\oslash\Gamma_M$. In
this section we give two explicit examples of such construction: the
1--dimensional C.Y. manifold described by a cubic polynomials in
CP(2):

$$
W={1\over 3} (y^3_1+y^3_2+y^3_3) -\psi y_1 y_2 y_3 =0\eqno(2.4)
$$

\noindent
and the 3--dimensional C.Y. manifold describe by a quintic polynomial
in CP(4):

$$
W={1\over 5} (y^5_1+y^5_2+y^5_3+y^5_4+y^5_5) -\psi
y_1 y_2 y_3 y_4 y_5 =0\eqno(2.5)
$$

\noindent
where $y^i, i=1,..., d+2$ are homogeneous coordinates in CP(d+1),
$d$ is the complex dimension  equal to 1 or 3 in our case and
$\psi$ is a single modulus parametrizing the complex structure
deformations of the hypersurface. Note that while in the case of
the torus the space of complex structure deformations is
one--dimensional, in the case of the quintic $W_0={1\over 5}
(y^5_1+y^5_2+y^5_3+y^5_4+y^5_5)$ there are 101 indipendent complex
structure deformations so that the moduli space is 101--dimensional.
The particular 1--dimensional  subspace described by the simple
deformation (2.5) is such that $1,y_1...y_5, (y_1...y_5)^2,
(y_1...y_5)^3$ close a subring of the chiral ring of all the
marginal operators associated to the deformations of the quintic.

Let us begin to study the case of the torus. From (2.4), using
known algorithms, (see ref. [15]),one obtains the following PFE's:

$$
{d\over d\psi}
{\omega_0\choose\omega_1}\ =\
\left(\matrix{
0                       &1 \cr
{\psi\over 1-\psi^3}    & {3\psi^2\over 1-\psi^3}\cr}\right)
{\omega_0\choose\omega_1}\eqno(2.6)
$$

\noindent
This can be traded for a single 2nd-order differential equation
for $\omega_0$

$$
\Bigl( {d^2\over d\psi^2} - {3\psi^2\over 1-\psi^3}
{d\over d\psi} - {\psi\over 1-\psi^3}\Bigr)\ \omega_0=0\eqno(2.7)
$$

\noindent
which exhibits four regular singular points at $\psi^3=1,
\psi=\infty$.

The monodromy group of this equation can be studied as follows.
First of all we note that it is sufficient to compute the
monodromy matrix $T_0$ around $\psi=1$. In fact the effect of a
closed loop around $\psi=\alpha$ and $\psi=\alpha^2$
$(\alpha = e^{2\pi i/3})$ can be computed from the monodromy matrix
$T_0$ around $\psi=1$ by conjugation with ${\cal A}$, where
${\cal A}$ represents the operation $\psi\rightarrow\alpha\psi$:

$$\eqalign{
T_1 =  &  {\cal A}^1 T_0 {\cal A}^{-1}\cr
T_2 =  &  {\cal A}^2 T_0 {\cal A}^{-2}\cr}\eqno(2.8)
$$

Furthermore a closed loop which encloses all the singular points,
including $\infty$, is contractible and therefore

$$
T_\infty T_2 T_1 T_0 = 1\quad\rightarrow\quad
T_\infty = (T_2 T_1 T_0)^{-1}\eqno(2.9)
$$

To compute $T_0$ it is convenient to perform the substitution
$z=\psi^3$ in the differential equation (2.7). We obtain

$$
\Bigl\{ 9z(1-z) {d^2\over dz^2} + (6-15z) {d\over dz} -1\Bigr\}
\omega=0\eqno(2.10)
$$

This is a hypergeometric equation of parameters $a=b=1/3, c=2/3$
and therefore a set of independent solutions around $z\equiv\psi^3=0$
is given by

$$\cases{
U_1  =  {\Gamma^2 (1/3)\over\Gamma(2/3)} F(1/3,1/3,2/3;\psi^3)\cr
U_2  =  {\Gamma^2 (2/3)\over\Gamma(4/3)} \psi
F(2/3,2/3,4/3;\psi^3)\cr}\eqno(2.11)
$$

\noindent
where $F(a,b,c;z)$ is the hypergeometric functions.

These 2 solutions can be continued around $\psi^3=1$ by known
formulae $^{16}$: one finds

$$\eqalign{
U_1 & =-\log(1-z) F (1/3,1/3,1;1-\psi^3)+B_1(1-\psi^3)\cr
U_2 & =-\log(1-z) F (1/3,1/3,1;1-\psi^3)+B_2(1-\psi^3)\cr}\eqno(2.12)
$$

\noindent
where $B_1$ and $B_2$ are regular series around $\psi^3=1$. (The
appearance of the logarithmic factor in (2.12) is traceable to the equality
of the roots of the indicial equation around $z\equiv\psi^3=1$). A closed
loop around $\psi=1$ gives

$$
{U_1\choose U_2}\rightarrow
{U'_1\choose U'_2}={U_1\choose U_2}
- 2\pi i\ F({1\over 3}, {1\over 3}, 1;
1-\psi^3)
{1\choose 1}\eqno(2.13)
$$

The Kummer relations $^{16}$ among hypergeometric functions allow us
to reexpress $F({1\over 3}, {1\over 3}, 1; 1-\psi^3)$ in terms of
the original basis $(U_1,U_2)$ around $\psi=0$,

$$
F\Bigl({1\over 3}, {1\over 3}, 1; 1-z\Bigr)=
{\Gamma({1\over 3})\over\Gamma^2({2\over 3})} F
\Bigl({1\over 3}, {1\over 3}, {2\over 3}; z\Bigr)
+ {\Gamma(-{1\over 3})\over\Gamma^2({1\over 3})} F
\Bigl( {2\over 3}, {2\over 3}, {4\over 3}; z\Bigr)\eqno(2.14)
$$

Therefore, using the relation  $\Gamma(z)\Gamma(1-z) =
{\pi\over \sin\pi z}$ one obtains

$$
{U'_1\choose U'_2}\quad = \quad
\left(\matrix{
1+i\ tg {2\pi\over 3}   & i\ tg {2\pi\over 3}\cr
i\ tg {2\pi\over 3}        & i-i\ tg {2\pi\over 3}\cr}\right)\quad
{U_1\choose U_2}\eqno(2.15)
$$

\noindent
that is the monodromy matrix around $\psi=1$ is

$$
T_0 =
\left(\matrix{
1-i\sqrt{3}   & i\sqrt{3}\cr
-i\sqrt{3}    & 1+i\sqrt{3}\cr}\right)\eqno(2.16)
$$

To find $T_1,T_2$ we need to represent ${\cal A}:\psi\rightarrow
\alpha\psi$ on $U_1, U_2$. From (2.6), (2.7) and (2.11) we
see that under $\psi\rightarrow\alpha\psi$ the differential operator
is invariant while

$$
{U_1\choose U_2} \rightarrow
\left(\matrix{
1 & 0\cr
0 & \alpha\cr}\right)
{U_1\choose U_2}\eqno(2.17)
$$

Since we are interested in the projective representation of the monodromy
group we may rescale our basis in such a way that $\det{\cal A}=1$
(note that $T_0$ already satisfies $\det T_0=1$).
Hence we have

$$
{\cal A}=
\left(\matrix{
\alpha^{-1/2} & 0 \cr
0             & \alpha^{1/2}\cr}\right)\eqno(2.18)
$$

\noindent
and from (2.8)

$$
T_1=
\left(\matrix{
1-i\sqrt{3} & \alpha^{-1}i\sqrt{3}\cr
-\alpha i\sqrt{3} & 1+i\sqrt{3}\cr}\right)\qquad
T_2=
\left(\matrix{
1-i\sqrt{3} & \alpha^{-2} i\sqrt{3}\cr
-\alpha^2 i\sqrt{3} &1+i\sqrt{3}\cr}\right)
\eqno(2.19)
$$

Let us now recall that the modular group is given by the group of
transformations on the variable $\psi$ which leaves the theory
invariant. The monodromy group $\Gamma_M$ of the PFE's must
therefore be a subgroup of the modular group. In our case the
modular group of the torus is known a priori to be $\Gamma=SL
(2;{\bf Z})$ and therefore it should be possible to perform a change
of basis on the periods $U_i$ such that the entries of the generators
$T_0, T_1, T_2$ are integer numbers. Actually it is known since the
last century that $\Gamma_M$ is isomorphic to $\Gamma(3)$, where
$\Gamma(3)$ is the group of matrices equivalent to the identity
modulo 3. The basis $({\cal F}_1, {\cal F}_2)$ where
$\Gamma_M\simeq\Gamma(3)$ is obtained by the following linear
transformation $^{17}$

$$
{{\cal F}_1\choose{\cal F}_2}\quad =\quad
{1\over 3(1+\alpha^{-1/2})}
\left(\matrix{
3\alpha^{1/2}    & -3\cr
1+\alpha^{1/2}   & \alpha^2-1\cr}\right)\quad
{U_1\choose U_2}\eqno(2.20)
$$

The transformed $\Gamma_M$ generators $\hat T_i$ take the
following form:

$$
\hat T_0 =\left(\matrix{1 &3\cr 0 &1\cr}\right)\quad;
\hat T_2 =\left(\matrix{-5 & 12\cr -3 &7\cr}\right)\quad;
\hat T_1 =\left(\matrix{-2 &3\cr -3 &4\cr}\right)
$$

$$
\hat T_\infty \equiv (\hat T_2 \hat T_1 \hat T_0)^{-1} =
\left(\matrix{1 &0\cr -3 &1\cr}\right)\eqno(2.21)
$$

The transformation ${\cal A}:\psi\rightarrow\alpha\psi$ is obviously
an invariance of $W=0$ (and of the  differential operator (2.17)) since
it can be undone by the coordinate transformation $y^i\rightarrow
\alpha^{-1/3} y^i$. Less evident is the invariance under the
transformation ${\cal B}$:

$$
{\cal B} : \psi' = - {\psi+2\over 1-\psi}\eqno(2.22)
$$

\noindent
which can be undone by the change of coordinates

$$
\left(\matrix{y'_1\cr y'_2\cr y'_3\cr} \right)\matrix
={i\over\sqrt{3}}
\left(\matrix { 1  &1  &1\cr
                1  &\alpha &\alpha^2\cr
                1  &\alpha^2 &\alpha\cr}
\right)
\left(\matrix {y_1\cr y_2\cr y_3\cr}\right)
\eqno(2.23)
$$

In the basis $({\cal F}_1, {\cal F}_2)$ $A$ and $B$  take the form

$$
\hat A =
\left(\matrix{ 1  &-3\cr 1 &-2\cr}\right)\ ;\
\hat B =
\left(\matrix{ 2 &1\cr 1 &1\cr}\right)
\eqno(2.24)
$$

We note that the $\Gamma_W$ generators $A, B$ satisfy
the relation $A^3=B^2=1$, $(AB)^3=1$ which are the defining
relations of the tetrahedral group $\Delta$. Indeed $\Gamma(3)$
is a normal subgroup of $\Gamma\equiv SL(2,{\bf Z})$, the modular
group of the torus ,and $\Gamma/\Gamma(3)\equiv\Gamma_W$.

We have thus verified that the relation $\Gamma/\Gamma_M\simeq
\Gamma_W$ actually holds in the case of the torus.

Let us now consider the case of the quintic (2.5) describing
a C.Y. 3--fold in CP(4). The PFE for the periods is given by
the following fourth--order\hfill\break
equation $^{4,5,15}$:

$$
{d^4V\over d\psi^4} -
{10\psi^4\over 1-\psi^5}
{d^3 V\over d\psi^3} - {25\psi^3\over 1-\psi^5}
{d^2 V\over d\psi^2} - {15\psi^2\over 1-\psi^5}
{dV\over d\psi} - {\psi\over 1-\psi^5}\ V=0\eqno(2.25)
$$

The four independent solutions of this equation represent the
four periods of the uniquely defined (3,0)--form $\Omega$ which
always exists on a C.Y. space; in our case it can be explicit
computed from the defining polynomial (2.5), but we do not need
its explicit form in the following. The periods are defined by
integrating $\Omega(y,\psi)$ on a basis $(\gamma_A, \gamma^B)$
of 3--cycles of $W$ satisfying

$$
\gamma^A\cap \gamma_B=-\gamma_B\cap\gamma^A=\delta^A_B
\qquad ; \qquad
\gamma^A\cap\gamma^B=\gamma_A\cap\gamma_B=0\eqno(2.26)
$$

The basis (2.26) is defined only up $Sp(4;{\bf Z})$ transformations,
which leave the intersection properties (2.26) invariant. We define

$$
V = (X^0, X^1, F_0, F_1) \equiv (X^A, F_A)\eqno(2.27)
$$

\noindent
where

$$
\int_{\gamma_A} \Omega = X^A(\psi)\ ; \
\int_{\gamma^A} \Omega = F_A(\psi)\eqno(2.28)
$$

Actually on a C.Y. 3--fold the four periods $V$ are not
functionally independent, but satisfy the relation (1.10)
where $F$ is a homogeneous function of degree two in
$X^0, X^1$. That has its counterpart in a functional relation
satisfied by the coefficients of the differential equation (2.25),
namely$^{18,19}$

$$
W_3 = a_1- {da_2\over d\psi} - {1\over 2}
{d^2 a_3\over d\psi^2} = 0\eqno(2.29)
$$

\noindent
where $a_i$ is the coefficients of  ${d^i\over dz^i}$
in the differential equation (2.25).

The reason why we have called $W_3$ the l.h.s. of eq. (2.29) is
that it coincides with the $W_3$ generator of the $W_4$--algebra
associated to a generic 4--th order linear differential operator
(2.25). Actually $W_3=0$ is an invariant statement since $W_3$
transforms as a covariant tensor of order 3 under
$\psi$--reparametrizations

$$
W'_3(\psi') = J^{-3} W_3 (\psi)\eqno(2.30)
$$

\noindent
where $J$ is the Jacobian of the transformation. The fact the
$W_3=0$ for the Picard--Fuchs equation of the quintic was first
noted by Lerche et al. [15]. Actually this constraint has been proven to
hold quite generally as a consequence of the fact that the
moduli space geometry is a special K\"ahler geometry $^{18,19}$. Indeed one
finds that in the one modulus case $W_3=0$ is equivalent to the
statement that the
associated $4\times 4$ linear system

$$
\Bigl( {d\over d\psi} + A\Bigr)\ \hat V=0\eqno(2.31)
$$

\noindent
where $\hat V\equiv (V,V',V'',V''')^t$, has a Drinfeld--Sokolov
connection $A$ which is gauge equivalent to a $Sp(4)$--connection
$^{20}$.

The gauge group acting on (2.31) is the group generated by the
strictly lower triangular matrices which leave invariant the top
component of $\hat V$ and therefore also the differential equation
(2.26).\footnote*{If one allows for rescaling of the solutions
of (2.25) then the gauge group can be extended to the Borel
subgroup of $GL(4)$.}
Note that in the gauge where $A$ is valued in the Lie
Algebra of $Sp(4)$ the matrix of solutions of (2.31) is an element
of $Sp(4)$

$$
\hat V =
\left(\matrix{
V_1  & V_2  & V_3 & V_4\cr
\star    & \star    & \star   & \star  \cr
\star    & \star    & \star   & \star  \cr
\star    & \star    & \star   & \star  \cr}\right)\quad
\in\ Sp (4)\eqno(2.32)
$$

In (2.32) the top row is the set of gauge invariant solutions of
(2.25). Of course the gauge (2.32) is not completely fixed since
arbitrary $Sp(4)$ gauge transformations on (2.31) keep the matrix
$\hat V$ in the symplectic gauge. Another consequence of the relation
of the PFE's with Special Geometry is that the coefficient $a_3$ of
${d^3\over dz^3}$ has always the following form $^{18}$:

$$
a_3 = - W^{-1} {dW^{-1}\over dz}\eqno(2.33)
$$

\noindent
where $W(\psi)$ is the Yukawa coupling of the effective $N=1$ theory
obtained by compactification on the given C.Y. 3--fold. As an example in
the quintic case by comparing (2.33) with (2.25) we find

$$
W(\psi) = {1\over 1-\psi^5}\eqno(2.34)
$$

The important thing to observe is that many of the properties that we
discussed in the one modulus case can be easily extended to the case of
$n$ moduli, where the PFE's are partial differential equations.
\noindent
Indeed, as it has been shown in ref.[19] in the ${n}$- moduli case the
linear system (2.31) can be generalized as follows:

$$ \Bigl(
{\partial\over\partial\psi^\alpha} +
A_\alpha (\psi)\Bigr) V (\psi) =0\eqno(2.35)
$$

\noindent
where ${V}$ is now a ${2n+2}$-dimensional vector and $A_\alpha$ is a
1-form matrix which in general is valued in the Lie Algebra of $GL (2n+2)$.
The set of $2n+2$ linearly independent vectors of (2.35) is now given
by a ${(2n+2)\times (2n+2)}$ matrix of the form

$$
V=\left(\matrix{
X^0    &   X^a   &F_a   &-F_0\cr
\star_{nx1}    & \star_{nxn}  &\star_{nxn}   &\star_{nx1}\cr
\star_{nx1}    &\star_{nxn}   &\star_{nxn}   &\star_{nx1}\cr
\star           &\star_{1xn}  &\star_{1xn}   &\star\cr}\right)
\eqno(2.36)
$$

\noindent
where $X^A = ( X^0,X^a)$, $F_A =( F_0,F_a)$, $a=1,...,n$ represent
the periods of the holomorphic $3$-form $\Omega$ defined in (2.28),
with the proviso that the index $A$ now takes the values $A =
0,1,...,n$.
\noindent
It can be shown that for C.Y. $3$-folds the higher order differential equations
equivalent to the linear system (2.35) is always a coupled set of partial
differential equations of order four.( Eq.(2.25)  is in fact an example of
this general rule in the case of a single modulus).This result can be shown
in general by using the Special Geometry identities,which are equivalent
to the Picard-Fuchs equations$^{19}$.Furthemore the same identities
also tell us that one can take advantage of the aforementioned gauge
invariance of the linear system  (2.35)  in such a way that the $A_\alpha$-
connection becomes valued in the Lie Algebra of $Sp (2n+2)$.Specifically
if we perform the gauge transformation

$$
V\rightarrow {\cal N}V \qquad;\qquad
{\cal N}=
\left(\matrix{
\star  &0_{1\times n} &0_{1\times n} &0\cr
\star_{n\times1}  &\star_{n\times n} &0_{n\times n}  &0_{n\times1}\cr
\star_{n\times1}  &\star_{n\times n}  &\star_{n\times n} &0_{n\times1}\cr
\star  &\star_{1\times n}  &\star_{1\times n}  &\times\cr}\right)
\eqno(2.37)
$$

$$
A'={\cal N}^{-1} A {\cal N} + {\cal N}^{-1} d{\cal N} \eqno(2.38)
$$

\noindent
where $\cal N$ belongs to the Borel subgroup of $GL (2n+2)$
then it is always possible to choose ${\cal N}$ in such a way that the linear
system be transformed in the "Special Geometry gauge",namely:

$$
(\partial_\alpha + A_\alpha) V =0\eqno(2.39)
$$

\noindent
where $A_\alpha=\pmb{\Gamma}_\alpha + {\bf C}_{\alpha}$, is given by:

$$
\pmb{\Gamma}_\alpha =
\left(\matrix{
-\partial_\alpha \hat K  &0  &0    &0   \cr
\noalign{\vskip 0.2truecm}
{\ 0}    &(\hat\Gamma_\alpha-\partial_\alpha\hat K{\bf 1})^\gamma_\beta &
0 &0\cr
\noalign{\vskip 0.2truecm}
0    &0    &(\partial_\alpha\hat K{\bf 1}-\hat\Gamma_\alpha)^\beta_\gamma &
0\cr
\noalign{\vskip 0.2truecm}
0    & 0   &0     &\partial_\alpha\hat K\cr}\right)
\eqno(2.40)
$$

$$
{\bf C}_\alpha =
\left(\matrix{
0     &\delta^\gamma_\alpha        &0         &0\cr
\noalign{\vskip 0.2truecm}
0     &0     &(W_\alpha)_{\gamma\beta}      &0\cr
\noalign{\vskip 0.2truecm}
0     &0            &0    &\delta^\beta_\alpha\cr
\noalign{\vskip 0.2truecm}
0     &0           &0         &0\cr}
\right)
\eqno(2.41)
$$
\vskip 10pt

\noindent
The hatted holomorphic connections $\partial_\alpha\hat K(z)$ and
$\hat\Gamma^\gamma_{\alpha\beta}(z)$ are given by

$$
\partial_\alpha\hat K = - \partial_\alpha\log X^0(z)\eqno(2.42)
$$

$$
\hat\Gamma^\gamma_{\alpha\beta} = e^{-1}_a (z)
\partial_\beta e^a_\alpha(z)\eqno(2.43)
$$

\noindent
where $e^a_\alpha=\partial_\alpha t^a(z), t^a(z)=X^a/X^0$.
They obviously vanish in the ``special coordinate" frame
$t^a=X^a/X^0$, $X^0=1$. Thus we see in particular that the
special coordinates $t^a$ coincide with the ``flat coordinates"
of the Landau--Ginzburg formulation of SCFT.
Note that in the special coordinate frame the connection ${\bf C_\alpha}\
\equiv A_\alpha$. The set of $\bf C_\alpha$-matrices is easily seen to satisfy
the following abelian,nilpotent subalgebra of $Sp (2n+2)$:

$$
{\bf C_\alpha}{\bf C_\beta}{\bf C_\gamma}{\bf C_\delta} = 0\eqno(2.44)
$$

$$
\Big[{\bf C_\alpha},{\bf C_\beta}\Big] = 0\eqno(2.45)
$$

\noindent
whose importance for the determination of the duality group of a general
C.Y.$3$-fold will be discussed later on. Note also that $\bf C_\alpha$ is
given in terms of the Yukawa coupling $W_{\alpha\beta\gamma}$ and that
${\bf C_\alpha}{\bf C_\beta}{\bf C_\gamma} = EW_{\alpha\beta\gamma}$ where $E$
is the $(2n+2)\times(2n+2)$ matrix with zeros everywhere except a $1$ in the
right upper corner.

The holomorphic form of
the special geometry equations give more insight into the
group--theoretical properties of the Picard Fuchs.equations. Indeed the
matrix $A$ of the linear system is not the most general one: in
the gauge (2.40-41) one easily verifies that actually it is valued in the
Lie algebra of $Sp(2n+2)\subset\ GL(2n+2)$. Indeed from (2.40-41) one
has

$$
A_\alpha Q = (A_\alpha Q)^t\eqno(2.46)
$$

\noindent
where $Q$ is the symplectic metric satisfying $Q^2=-1, Q^t=-Q$

$$
Q=\left(\matrix{
     &      &            &1\cr
     &      &-{\bf 1}_n  & \cr
     &{\bf 1}_n   &      & \cr
-1   &      &            & \cr}\right)\eqno(2.47)
$$

This in turn implies that the period matrix $V$ is valued in the
$Sp(2n+2)$ group. In particular the top row $V$ of the gauge
invariant solutions is defined only up to symplectic transformations
$M$

$$
V' = VM, \quad\quad M\in\ Sp(2n+2)\eqno(2.48)
$$

These transformations leave invariant the K\"ahler potential defined
in (1.8) since it can be rewritten as $K=-\log(V(-iQ)V^\dagger)$.
For more details on the many moduli case see Ceresole et al.$^{19}$.

Let
us now come back to the discussion of a single modulus.
Our aim is to discuss the modular group $\Gamma$ of the moduli space of
the quintic polynomial (2.5).

Let us first consider the duality group $\Gamma_W$ of the defining
polynomial equation $W=0$ given by (2.5). It is obvious that
${\cal A}:\psi\rightarrow\alpha\psi$, where $\alpha\equiv e^{2\pi i/5}$,
is a symmetry of $W=0$ since it can be undone by a rescaling of the CP
(4) homogeneous coordinates:
$(y_1, y_2, y_3, y_4, y_5) \rightarrow $ $(\alpha^{-1}y_1,
y_2,y_3,y_4,y_5)$. Obviously ${\cal A}^5=1$ and
this excludes a priori the possibility that we can represent the modular
group as a subgroup of $SL(2;{\bf Z})$ acting projectively on a function
of $\psi$, since $SL(2,{\bf Z})$ does not possess elements of order 5.
Since there are apparently no other $\psi$--transformations which can be
undone by linear transformations of the $y_i '$s, $\Gamma_W$ is simply
the cyclic group ${\bf Z}_5$. According to our previous discussion to
reconstruct the full modular group we must now compute the monodromy
group $\Gamma_M$ of eq. (2.25).

$\Gamma_M$ will be represented by $4\times 4$ matrices on the four
periods $(V_1, V_2, V_3, V_4)$ solutions of (2.25). The same is true
for $\Gamma_W$ since ${\cal A}:\psi\rightarrow\alpha\psi$ leaves
invariant the differential operator of eq. (2.25) and therefore induces
just a linear combinations of the periods. Furthermore by using the
gauge (2.32),(or,in particular,the Special Geometry gauge),we may represent
$\Gamma_M$
 and $\Gamma_W$ by
$Sp(4;{\bf Z})$--matrices.

Let us first compute $\Gamma_M$. We sketch briefly the procedure, for
further details see ref. [4].

The differential equation (2.25) is a Fuchsian equation with
regular singular points at $\psi=\alpha^k$, $(k=0,1,...,4)$,
$\alpha=e^{2\pi i/5}$, and $\psi=\infty$. As in the case of the
torus it is sufficient to study the monodromy matrix $T_0$ around
$\psi=1$, since around $\psi=\alpha^k$ the corresponding monodromy
matrices $T_k, k=1,2,3,4$ are given by

$$
T_0\rightarrow T_k = A^k T_0 A^{-k}\eqno(2.49)
$$

\noindent
where $A$ represents $\psi\rightarrow\alpha\psi$. The monodromy around
$\psi=\infty$ depends on the other generators around $\psi=\infty$
through the relations $T_\infty T_4 T_3 T_2 T_1 T_0=1$.

As in the torus case it is convenient to transform eq. (2.25) into a
generalized hypergeometric equation through the substitution
$z=\psi^{-5}$. We obtain

$$\eqalign{
\Bigl\{ {d^4\over dz^4} & - {2(4z-3)\over z(1-z)} {d^3\over dz^3} -
{72z-35\over 5z^2(1-z)} {d^2\over dz^2} -
{24z-5\over 5z^3(1-z)} {d\over dz} - \cr
& - {24\over 625 z^3 (1-z)} \Bigr\} V(z)=0\cr}\eqno(2.50)
$$

\noindent
which has singular Fuchsian points at $z=0,1,\infty$ with associated
Riemann P--symbol

$$ P
\left\{ \matrix{
0    &\infty    &1    & \cr
0    & 1/5      &0    & \cr
0    & 2/5      &1    & ;\psi^{-5}\cr
0    & 3/5      &2    &   \cr
0    & 4/5      &1    &   \cr} \right\}
\eqno(2.51)
$$

We notice that in the variable $z=\psi^{-5}$ we have introduced a new
singular point around $\psi=0$ so that the monodromy around $z=\infty$
corresponds exactly to the representation of $\psi\rightarrow\alpha
\psi$ on the periods. In other words the duality generator $A$ becomes part
of the monodromy generators of the new equation (2.50).\footnote*{The same
observation can be done in the case of the torus previously discussed.}
A solution of
(2.50) around $z=0$ $(\psi=\infty)$ is given by

$$
\omega_0(\psi) =\ _4 F_3
\Bigl( {1\over 5}, {2\over 5}, {3\over 5}, {4\over 5}; 1,1,1;
\psi^{-5}\Bigr)\eqno(2.52)
$$

In order to represent $A$ in a simple way we may construct a basis of
solutions around $\psi=0$ as follows$^{4}$. We first continue $\omega_0(\psi)$
around $\psi=0$ by using a Barnes--type integral representation and we
obtain:

$$
\omega_0(\psi) = - {1\over 5\cdot 16 \pi^4}
\sum^\infty_{n=0} {\Gamma^5({n\over 5})\over\Gamma(n)}
(\alpha^n-1)^4 (5\psi)^n\qquad (|\psi< 1)\eqno(2.53)
$$

Then we recall that $\psi\rightarrow\alpha\psi$ leaves the differential
operator (2.25) invariant so that

$$
\omega_j(\psi) \doteq \omega_0 (\alpha^j\psi)\qquad\qquad
j=0,1,2,3,4\eqno(2.54)
$$

\noindent
are also solutions of (2.50). The five functions $\omega_j$ are
subject to the linear relations $\sum^4_{n=0} \omega_j=0$, as it
follows from their explicit expression by the power series (2.53)
and the analogous ones
derived from (2.54).

If we take $\omega_0, \omega_1, \omega_2, \omega_4$ as a basis of
solutions around $\psi=0$ it follows immediately that $\psi
\rightarrow\alpha\psi$ is represented on $(\omega_2, \omega_1,
\omega_0, \omega_4)^t$ as follows

$$
A=
\left( \matrix{
-1   & -1     &-1    &-1\cr
1    & 0      & 0    &0 \cr
0    & 1      & 0    & 0\cr
0    & 0      & 1    & 0\cr}\right)\eqno(2.55)
$$

Next one examines the monodromy $T_0$ around $\psi=1$. For this purpose one
observes that since $z=1$ has the double root $\rho=1$ for the indicial
equation, the continuation of the series $\omega_j(\psi),|\psi|<1,$ to
the neighbourhood $|\psi-1|<1$ will contain logarithms. Indeed one can
write

$$
\omega_j(\psi) = {1\over 2\pi i} c_j \tilde\omega(\psi)
\log (\psi-1) + reg.\eqno(2.56)
$$

\noindent
where $\tilde\omega$ is a linear combination of regular solutions
around $\psi=1$. It turns out that

$$
\tilde\omega(\psi) = - {1\over c_1}
(\omega_1(\psi)-\omega_0(\psi))
={1\over c_1} (\psi-1) + O (\psi-1)^2)\eqno(2.57)
$$

\noindent
It follows

$$\eqalign{
{d\omega_j\over d\psi} & = {1\over 2\pi i} c_j
\Bigl( {d\tilde\omega\over d\psi} \log(\psi-1)+
\tilde\omega(\psi) {1\over\psi-1} + ......\Bigr)\cr
& =_{\psi\rightarrow 1} {1\over 2\pi i} {c_j\over c_1} \log
(\psi-1)+\cdots\cr}\eqno(2.58)
$$

\noindent
We see that in order to compute the monodromy coefficients $c_j$
one has to compute the asymptotic behaviour of ${d\omega_j\over
d\psi}$ and look at the coefficient of log$(\psi-1)$. Using the
series expansion for $\omega_j$ derived from (2.39) and (2.40)
one finds (see [4] for details)

$$
c_j = (1,1,-4,6,-4)\eqno(2.59)
$$

\noindent
{}From eqs. (2.56-59), one easily finds that the monodromy
matrix around $\psi=1$ acting on the basis $(\omega_2, \omega_1,
\omega_0, \omega_4)^t$ is given by

$$
T_0 = \left( \matrix{
1  &4  &-4  &0  \cr
0  &0  &1   &0  \cr
0  &-1  &2   &0  \cr
0  &4   &-4   &1\cr}\right)\eqno(2.60)
$$

\noindent
The matrices $A$ and $T_0$ given by eqs. (2.55) and (2.60) are
integer valued, but not symplectic, since the $\omega_j$--basis
is not a symplectic basis. According to our previous discussion
there must exist a matrix $m$ such that

$$
\hat T_0 = m T_0 m^{-1}\quad;\quad
\hat A = m A m^{-1}\eqno(2.61)
$$

\noindent
are not only integer--valued but also symplectic.

A solution for $m$ has been found in ref. [4] which is unique
up to $Sp(4;{\bf Z})$ transformations.

We choose the following solution:

$$
\hat A=\left( \matrix{
1  &-1  &-5  &3\cr
0   &1  &8  &5\cr
1  &-1 &-4  &3\cr
0   &0  &1  &1\cr}\right)\quad;\quad
\hat T_0 =\left( \matrix{
1   &0   &0   &0\cr
0   &1   &0   &0\cr
-1  &0   &1   &0\cr
0   &0   &0   &1\cr}\right)\eqno(2.62)
$$

\noindent
which act on the right of the row vector (2.27).
The other monodromy generators $\hat T_k$, $\hat T_\infty$
around $\psi=\alpha^k$ and $\psi=\infty$ are finally computed from
eqs. (2.49) and $T_\infty=(T_4 T_3 T_2 T_1 T_0)^{-1}$.

The conclusion
is the following: the duality group $\Gamma$ of the moduli space of
the C.Y. 3--fold (2.5) can be given a $4\times 4$ representation on
the integer valued and symplectic basis of the periods. $\Gamma$ is
a subgroup of $Sp(4;{\bf Z})$ generated by the matrices $\hat A$, $\hat T_k
 (k=0,1,2,3,4)$, where $\hat A$ is a representation of the ${\bf Z}_5$
subgroup of $Sp(4;{\bf Z})$ which leaves $W=0$ invariant and the
$\hat T_k$'s generate the monodromy group of the PFE (2.25).

We have already observed that the group $\Gamma$ cannot be a subgroup of
$SL(2,{\bf Z})$ since $SL(2,{\bf Z})$ does not contain elements
of order 5. We may however represent  $\Gamma$ as a subgroup of
$SL(2,{\bf R})$ since ${\bf Z}_5 \in SL(2,{\bf R})$. To find the representation
we need  a variable $\gamma(\psi)$ such that ${\cal A}:\psi
\rightarrow\alpha\psi$ and the transport around $\psi=1$ are
represented as $PSL(2,{\bf R})$ transformations on $\gamma(\psi)$.
The determination of $\gamma(\psi)$
and the associated $2\times 2$ representation of
$\Gamma\in SL(2,{\bf R})$
has been given in ref. [4] by requiring that $\gamma(\psi)$ be a
modular parameter on which $A$ and $T_0$ act as transformations of
order five and infinity, respectively, on the upper $\gamma$--plane.
Standard formulae of the theory of the automorphic functions then
determine $\gamma(\psi)$. We give here a different, but closely
related derivation, which is based only on the structure of the PFE
(2.25).

Let us observe that if we denote by $V(\psi(\gamma))$ the
four--dimensional row vector of the periods as a function of $\gamma$,
then we must have:

$$
V\Bigl(\psi\Bigl( {a_i\gamma+b_i\over c_i\gamma+d_i}\Bigr)\Bigr)=V
(\psi(\gamma)) \Gamma_i\eqno(2.63)
$$

\noindent
where $\Gamma_i$ is any of the matrices $\hat A, \hat T_k$ and
$S_i \equiv {a_i\gamma+b_i\over c_i\gamma+d_i}$ the corresponding
2--dimensional action on $\gamma$. If $V$ is required to be a
uniform function of $\gamma$, then $\psi=\psi(\gamma)$ must be
uniform and such that the entire $\psi^5$--plane is mapped into a
fundamental region of the $\gamma$--plane for the group
$\Gamma\equiv\{S_i\}$. That amounts to say that $\gamma$ is a
modular variable and $\psi$ is an automorphic function of $\gamma$
with respect to $\Gamma$. There is a general procedure to construct
the uniformizing variable $\gamma$ directly from the PFE for $V$.
It consists in associating to the main differential equation,
eq. (2.50) in our case, a second order differential equation with
the same singular points $(z=0,1,\infty$ in our case) and with
exponents determined as follows.
\noindent
If all the integrals of the main equation are regular around the
given singularity (no two roots of the indicial equation differ
by integers) and if all the roots are commensurable quantities
multiple of $1/k$ ($k$ integer), then the difference of the
roots of the indicial equations of the associated 2nd--order
equation is taken equal to $1/k$. In all the other cases the
difference of roots is taken equal to zero. The uniformizing
variable $\gamma$ is then given by the ratio of two solutions
of the associated 2nd--order equation. Let us see how this
works in our case. To adhere to the same notations as in [4]
we perform the substitution $z\rightarrow {1\over z}$ in the
equation (2.50). The P--Riemann symbol (2.37) becomes transformed into

$$P
\left( \matrix{
0   &\infty    &1     &\cr
1/5 & 0        &0     &\cr
2/5 & 0        &1     &;\psi^5\cr
3/5 & 0        &2     &\cr
4/5 & 0        &1     &\cr}\right)\eqno(2.64)
$$

\noindent
{}From (2.64) we see that at $z^{-1}\equiv\psi^5=0$ all the roots
are multiple of ${1\over k} \equiv {1\over 5}$ and do not differ
by integers. At $z^{-1}=\infty$ and $z^{-1}=1$ instead we have at
least two coincident roots. Therefore calling $\lambda,\mu,\nu$ the
differences of the roots of the indicial equation for the associated
2nd--order equation we have

$$
\lambda=1/5\qquad ;\qquad \mu=\nu=0\eqno(2.65)
$$

Given the exponent we can immediately write down the associated
2nd--order equation, which, having regular singular points at
$z=0,1,\infty$ is a hypergeometric equation of parameters

$$
a={1\over 2}(1-\lambda-\mu+\nu)= {2\over 5};\
b= {1\over 2} (1-\lambda-\mu-\nu) = {2\over 5};\
c=1-\lambda = {2\over 5}\eqno(2.66)
$$

\noindent
that is

$$
z(1-z){\cal F}''+\Bigl( {4\over 5} - {7\over 5} z\Bigr) {\cal F}'
- {4\over 25} {\cal F}=0\eqno(2.67)
$$

\noindent
The uniformizing variable is then given by

$$
\gamma= {{\cal F}_1\over{\cal F}_2}\eqno(2.68)
$$

\noindent
where

$$
{\cal F}_1 \equiv {\Gamma^2({2\over 5})\over\Gamma({4\over 5})}
F (2/5, 2/5, 4/5; \psi^5)\eqno(2.69)
$$

$$
{\cal F}_2 \equiv {\Gamma^2({3\over 5})\over\Gamma({6\over 5})}
F (3/5, 3/5, 6/5; \psi^5)\eqno(2.70)
$$

\noindent
are two linearly independent solutions of (2.67). From the theory
of the automorphic functions we know that $\gamma$ maps the
$\psi^5$--plane onto a couple of adjacent triangles inside the
circle $|\gamma|^2=1$ with internal angles $(0,0,\pi/5)$; they
constitute a fundamental region for the projective action of the
modular group $\Gamma$, and the inverse function $\psi=\psi(\gamma)$
is automorphic with respect to $\Gamma$.

It is now easy to derive the explicit representation of $A$ and
$T_0$ as a subgroup of $SL(2,{\bf R})$ on ${\cal F}_1, {\cal F}_2$:
one has simply to perform the study of the monodromy group of the
differential equation (2.53) in exactly the same way as we did it
in the case of the torus, eq. (2.10). One obtains in this case, with
completely analogous calculations:

$$ A=
\left( \matrix{
e^{-i\pi/5}   & 0\cr
0          & e^{i\pi/5}\cr}\right)\quad;\quad
T_0=
\left( \matrix{
1-i tg {2\pi\over 5}    & i tg {2\pi\over 5} \cr
-i tg {2\pi\over 5}     & 1+i tg {2\pi\over 5}\cr}\right)\eqno(2.71)
$$

\noindent
and

$$
T_k=A^k T_0 A^{-k}\quad
(k=1,2,3,4);\quad
T_\infty=(T_4 T_3 T_2 T_1 T_0)^{-1}\equiv(AT)^{-5}\eqno(2.72)
$$

\noindent
Note that the matrices quoted in the Candelas et al. $^4$ are related
to those given in (2.57) by the change of basis

$$
{Z_1\choose Z_2} =
\left(\matrix{ i   & -i \alpha^2\cr
               1   & -\alpha^2\cr}\right)
{{\cal F}_1\choose{\cal F}_2}
\eqno(2.73)
$$

\noindent
which map the interior of the circle $|\gamma|^2=1$ into the upper
half--plane $Im \gamma > 0$.

Till now we have considered the representation of the modular group
$\Gamma$ on the periods as $4\times 4$ $Sp(4;{\bf Z})$--valued matrices,
or as $SL(2;{\bf R})$--matrices acting projectively on the unformizing
variable $\gamma$. There is however another important variable, the
special variable $t$, in terms of which we may give an interesting
representation of $\Gamma$ which we now discuss.

To introduce it let us recall that in general, for any number $n$
of moduli, a symplectic transformation on the period vector
$V=(X^A, F_A)\equiv(X^0, X^a, F_0, F_a)$, $a=1,...,n,$ induces a
reparametrization on the special coordinates $t^a\equiv {X^a\over X^0}$.
Indeed from

$$\eqalign{
\tilde V & = V M\cr
M  & =
\left( \matrix{ A &C\cr B  &D\cr}\right)
\quad \in Sp (2n+2,{\bf R})\cr}\eqno(2.74)
$$

\noindent
where $A,B,C,D$ are $(n+1)\times(n+1)$ matrices obeying

$$\eqalign{
A^t B & = B^t A\cr
C^t D & = D^t C\cr
A^t D & - B^tC = {\bf 1}\cr}\eqno(2.75)
$$

\noindent
we readily obtain $^{19}$

$$
\tilde t^a = {\tilde X^a\over \tilde X^0} =
{A^a_B X^B + B^{aB} F_B\over A^0_B X^B + B^{0B} F_B}\eqno(2.76)
$$

\noindent
Recalling that $F_A = {\partial F\over \partial X^A}$ where $F(X)$
is a homogeneous function of degree two we have

$$\eqalign{
F(X^A) & = (X^0)^2{\cal F}(t^a)\cr
F_0 (X^A) & \equiv {\partial F\over\partial X^0} =
X^0 [2{\cal F}(t^a)-t^a\partial_a{\cal F}(t^a)]\cr
F_a(X^A)& \equiv {\partial F\over\partial X^a} = X^0
\partial_a{\cal F}(t^a)\cr}\eqno(2.77)
$$

\noindent
and substituting in (2.76) we find

$$
\tilde t^a =
{ A^a_b t^b + A^a_0 + B^{ab}{\cal F}_b + B^{a0} (2{\cal F}-t^b
{\cal F}_b)\over
A^0_b t^b + A^0_0 + B^{ab} {\cal F}_b + B^{00}
(2{\cal F}-t^b {\cal F}_b)}\eqno(2.78)
$$

\noindent
where ${\cal F}_a\equiv\partial_a{\cal F}$. If we now restrict the generic
$Sp(2n+2,{\bf R})$ matrix $M$ to belong to $\Gamma\subset Sp(2n+2,{\bf Z})$ we
obtain the representation of $\Gamma$ on the variables $t^a$. In particular
the subgroup of $Sp(2n+2;{\bf R})$ consisting of matrices of the form

$$
\left( \matrix{ A &C\cr 0 &D\cr}\right)\qquad
{D=(A^t)^{-1}\atop C=AC^t(A^t)^{-1}}\eqno(2.79)
$$

\noindent
act on the $t^{a'}$s as a group of linear fractional transformations. It
is compelling to assume that the subgroup ${\cal T}\subset PSL(2,{\bf Z})$
of discrete integer translations

$$
t^a\rightarrow t^a + n^a\quad ;\quad n^a \in {\bf Z}^n\eqno(2.80)
$$

\noindent
is always contained in $\Gamma$. Indeed the symmetry (2.80) has its stringy
origin
in the Wess--Zumino term for the $B_{ij}$ axion 2--form in the
$\sigma$--model.

Indeed the $\sigma$--model term

$$
\int_{W.S.} d^2\sigma \partial_\alpha Y^i\partial_\beta
\bar Y^{\bar j} B_{i\bar j} \epsilon^{\alpha\beta}=
\int_T B_{i\bar j} dY^i \wedge d\bar Y^{\bar j}\eqno(2.81)
$$

\noindent
where $T$ is the image of the world-sheet (W.S.) in the C.Y. 3--fold,
is topological in nature and the $B_{ij}$ are analogous to the
$\theta$--parameters of Q.C.D. If the complexified
K\"ahler (1,1)--form is parametrized as

$$
g_{i\bar j} +i B_{i\bar j} =-i
\sum^n_{a=1} t^a L^a_{ij}\eqno(2.82)
$$

\noindent
where the $L^a_{i\bar j}$ are a basis of the (1,1)--cohomology,
a shift $t^a\rightarrow t^a+n^a,n^a
\in{\bf Z}$ induces a topologically non
trivial mapping from the world sheet to the C.Y. and corresponds to an
instanton
correction to the $\sigma$--model perturbative result. Such integral
shift is an invariance of the quantum action.

In the one modulus case the existence of the translation symmetry
$t\rightarrow t+1$ has been verified in the case of the quintic by
Candelas et al.$^4$ and proven for a large class of 3--fold by Morrison
$^5$.
Let us verify it for the quintic. From the explicit form of the two
generator $A,T_0$ given by eq. (2.48) we find that on the
$(X^0, X^1, F_0, F_1)$--basis

$$
({T_0}A)^{-1} =
\left(\matrix{
1  &1  &5  &-8\cr
0  &1 &-3  &-5\cr
0  &0  &1   &0\cr
0  &0 &-1   &1\cr}\right)\eqno(2.83)
$$

\noindent
so that

$$
\tilde t\equiv {\tilde X^1\over \tilde X^0} =
{X^1\over X^0} +1 \equiv t+1\eqno(2.84)
$$

The $t$--transformations realized by A and $T_0$ are instead
non linear:

$$
A: \tilde t = {X^1-(X^0+F^0)\over X^0+F^0}=
{t-1-(2{\cal F}-t{\cal F}')\over 2{\cal F}-t{\cal F}'}\eqno(2.85)
$$

$$
T_0 : \tilde t = {X^1\over X^0-F_0} =
{t\over 1-2{\cal F} + t{\cal F}'}\eqno(2.86)
$$

\noindent
We note that while $({T_0}A)^{-1}$ corresponds to a circuit around
$z=0$, the monodromy around $\psi=\infty$ is represented by
$({ T_0}A)^{-5}$ so that

$$
({ T_0}A)^{-5} :\qquad \tilde t \rightarrow t+5\eqno(2.87)
$$

\noindent
The transformations (2.87) and (2.85) or (2.86) generate the whole
modular group on the $t$--modulus.

Coming back to the case of $n$ moduli $t^a$, we may also write
down the transformation law of the prepotential $F(X)$ in the general
case.
Indeed recalling the homogeneity relation $2F(X)=X^A F_A$ one finds for
a generic $Sp(4)$--transformations (2.74) $^{21}$:

$$\eqalign{
2\tilde F(\tilde X) & = (X^A, F_A)
\left(\matrix{AC^t  & AD^t\cr BC^t  &BD^t\cr}\right)
{X^A\choose F_A}\cr
= 2F(X) &+ 2 F_A(BC^t)^A_B X^B + X^A (AC^t)_{AB} X^B
+ F_A(BD^t)^{AB} F_B\cr}\eqno(2.88)
$$

\noindent
where we have used the conditions of symplecticity of the transposed
matrix $M^t$ in order to reconstruct $F(X)$ on the r.h.s. of (2.88).

If $M\subset\Gamma$, then $\tilde F=F$ since a modular transformation
is a discrete isometry. In particular for a translation (2.74) gives

$$
F(X^B A_B^A)=F(X^A) + X^A (AC^t)_{AB} X^B\eqno(2.89)
$$

\noindent
since $B=0$ and $A=\left(\matrix{1 & n^a\cr 0 &\delta^a_b\cr}\right)$.
In the $t^a$--variables eq. (2.89) becomes

$$
{\cal F}(t^a+n^a)={\cal F}(t^a)+(AC^t)_{ab} t^a t^b + 2
(AC^t)_{0b} t^b + (AC^t)_{00}\eqno(2.90)
$$

Thus $F$ (or ${\cal F}$) is periodic in the $X^A$ (or $t^a$) up to
quadratic additions. In particular the Yukawa coupling $W_{abc}=
{\partial^3{\cal F}\over \partial t^a \partial t^b \partial t^c}$
is periodic, $W_{abc}(t^a+n^a)=W_{abc}(t^a)$ and can be expanded in
a multiple Fourier series:

$$
W_{abc} (t^a) = \sum_{\vec m\in{\bf Z}^n} d_{abc}
(\vec m) e^{2\pi i\vec m\cdot\vec t}\eqno(2.91)
$$

\noindent
or, by changing variables $q_i=e^{2\pi it_i}$, $i=1,...,n$

$$
W_{abc} (q_i) = \sum_{\vec m\in{\bf Z}^n} d_{abc} (\vec m)
\Pi^n_{i=1} q^{m_i}_i\eqno(2.92)
$$

Note that $q_i=0$ means $t_i\rightarrow\ i\infty$, that is large
radius limit. Therefore the series (2.92) has a constant term
correspondig to $\vec m=0$ which gives the constant perturbative
Yukawa coupling $d_{abc}(0)$ while the $\vec m\not =0$ terms classify
the instanton corrections to the perturbative result.

In the quintic case the number $d_{111}(m)$ have been identified
with the number of rational curves of degree $n$ existing in the
quintic$^4$. The basic observation now is that the $n$ abelian
elements of the quantum duality group, related to the discrete
Peccei--Quinn symmetry (2.84) are $n$ symplectic $Sp(2n+2)$--matrices
entirely determined by the intersection numbers $d_{abc}(0)$, which
are topological classical objects computable in the large ``radius"
limit, $t^a\rightarrow\ i\infty$. This can be explicitly shown in the one
modulus case using the monodromy properties of the PFE around $q_i=0$.
It has been verified in a two--moduli case by Candelas et al. $^{22}$.

Actually in the one--modulus case the monodromy matrices can be
computed from the $A_\alpha$--connection of the linear system (2.31).
Using the change of variable $\psi\equiv t\rightarrow {1\over 2\pi i}
\log\ q$ the linear system becomes:

$$
\Bigl[ q {\partial\over\partial q} + {1\over 2\pi i} A(q)\Bigr]
V(q)=0 \eqno(2.93)
$$

\noindent
The monodromy generator ${\bf T}$ around $q=0$ is then given by
(see Morrison, ref. [5]):

$$
{\bf T} = exp\ A(q=0)\eqno(2.94)
$$

\noindent
and it has the property $({\bf T}-{\bf 1})^4=0$ which corresponds to
the maximal nilpotency dictated by the order of the differential
equation.

If we now recall the structure of the A--connection in special
coordinates (see (2.39-2.41), with $\pmb{\Gamma}=0)$:

$$
{d\over dt} V = {\bf C}(t) V \equiv
\left(\matrix{
0   &1   &0       &0\cr
0   &0   &W_{ttt} &0\cr
0   &0   &0       &1\cr
0   &0   &0       &0\cr}\right) V\eqno(2.95)
$$

\noindent
we see that for any $W$

$$
(exp\ {\bf C} - {\bf 1})^4 = 0\eqno(2.96)
$$

\noindent
as a consequence of ${\bf C^4}=0$ (eq. 2.44).

Therefore the symmetry ${(T_0)A}^{-1}:t\rightarrow t+1$, is
identified,up to a symplectic transformation, with $exp{\bf C} (t=i\infty)$,
where  $W(i\infty)\equiv
d_{111}$.

In the $n$--moduli case the $n$--monodromy generators
${\bf T}_i=exp {\bf C}_i(i\infty)={\bf 1}+{\bf L}_i$ satisfy
the following relations:

$$\eqalign{
[{\bf L}_i, {\bf L}_j] & = 0\cr
{\bf L}_i {\bf L}_j {\bf L}_k & = d_{ijk} E\cr
{\bf L}_i {\bf L}_j {\bf L}_k {\bf L}_l & = 0\cr}\eqno(2.97)
$$

\noindent
which follow from eqs. (2.44), (2.45) and the related ensuing
observation.

\vskip 20pt
\noindent
{\bf 3.\ Conclusions}
\vskip 10pt
\noindent
In theses lectures we have described at some length the mathematical
origin of target space duality symmetry, its relation to $N=2$
Superconformal field theories, algebraic geometry and effective actions
described by Supergravity Lagrangians.

The important discovery of mirror symmetry in the case of Calabi--Yau
string compactifications allows one to compute instanton
non--perturbative
$\sigma$--model corrections to the effective action purely in terms
of algebraic geometrical methods supplemented with the notion
of special geometry. When applied to space--time fermions,
target space duality, which is a discrete isometry of
the moduli space,
is seen as a compensating (field dependent) holonomy
transformation. Since fermions live in a complex representation of the
holonomy group $H\in U(N)$ ($N$ is the dimension of the K\"ahler
manifold), target space duality transformations induce $\sigma$--model
types of anomalies and in particular mixed $U(1)$ gauge anomalies
$^{14}$.
In supersymmetric effective gauge theories derived from strings the anomaly
cancellation mechanism which takes place originated from a 4D
Green--Schwarz mechanism for the 4D antisymmetric tensor $b_{\mu\nu}$
as well as by the existence of local Wess--Zumino terms which involve
automorphic functions of the target space modular group. By
supersymmetry they induce a moduli dependence on the gauge coupling
constant, $\Delta_a$, with important effects for the running toward
the string unification scale:

$$
{1\over g^2_a(\mu)} = k_a {1\over g^2_{string}} +
{b_a\over 16\pi^2}\ \log {M^2_{string}\over \mu^2} +
\Delta_a\eqno(3.1)
$$

\noindent
where $k_a$ are the levels of the Kac--Moody algebra for the gauge group
factors $G_a$, ${ g^{-2}_{string}}$ is the v.e.v. of the dilaton field,
$M^2_{string}$ is the string mass scale, $M^2_{string}=a\alpha'^{-1}$
($a$ numerical constant), and $b_a$ is the field theoretic 1--loop $\beta$--fun
ction
for the gauge group factor $G_a$

$$
b_a = - 3T (G_a) + \sum_R T(R)\eqno(3.2)
$$

\noindent
( $T(G_a)$ is the Casimir of the adjoint and $T(R)$ are the quadratic
Casimirs of the representations $R$ of chiral multiplets).

For a C.Y. manifold with $G=E_6\oplus E_8$ the $M_{GUT}$--scale occurs
then at the value

$$
M^2_X=M^2_{string} \Bigl( {Y_8\over Y_6}\Bigr)^{{
1\over b_8-b_6}}\qquad (\Delta_a={1\over 16\pi^2}\log Y_a)\eqno(3.3)
$$

\noindent
where $Y_a$ are automorphic functions of the target space duality
group $\Gamma$. The non--harmonic part of $\Delta_a$ can be entirely
computed from quantities in special geometry constructed out of the
holomorphic period vector $(X^A, F_A)$, namely the norm
$X^A\bar F_A-\bar X^A F_A$ and $\det \Bigl( {\partial^2 F\over
\partial X^A\partial X^B}\Bigr)$.

Threshold effects at the string scale, due to moduli dependence,
may play a crucial role for the discussion of supersymmetry breaking
when non--perturbative stringy effects, such as gaugino condensation,
are suitably incorporated.

\vskip 20pt
\noindent
{\bf Acknowledgements}
\vskip 10pt
\noindent
We would like to thank P.Candelas,A. Ceresole,P.Fre`,W.Lerche,J.Louis and
P.Soriani for enlightening discussion.
\vskip 20pt
{\bf References}
\vskip 10pt
\item{1.} P. Candelas, G. Horowitz, A. Strominger and A. Witten,
{\sl Nucl. Phys.} B268 (1985) 46.
See also  M. Green, J. Schwarz
   and E. Witten, {\sl Superstring Theory} (Cambridge, Univ. Press,
   1987).

\item{2.} E. Martinez, {\sl Phys. Lett.} B217 (1989) 431;
\item{} C. Vafa and N.P. Warner, {\sl Phys. Lett.}  B218 (1989) 51;
\item{} W. Lerche, C. Vafa and N. Warner, {\sl Nucl. Phys.}
    B324 (1989) 427;
\item{} D. Gepner, {\sl Phys. Lett.}  B222 (1989) 207;
\item{} P. Howe and P. West, {\sl Phys. Lett.}  B223 (1989) 377;
\item{} S. Cecotti, L. Girardello and A. Pasquinucci, {\sl Nucl. Phys.}
 B328 (1989) 701; {\sl Int. Jou. Mod. Phys.} A6 (1991) 2427;
\item{} C. Vafa, {\sl Int. J. Mod. Phys.}  A6 (1991) 2829;
\item{} K. Intrilligator and C. Vafa, {\sl Nucl. Phys.}  B339
   (1990) 95;
\item{} C. Vafa, {\sl Mod. Phys. Lett.}  A4 1615 and {\sl Mod. Phys.
   Lett.} A4 (1989) 1169;
\item{} S. Cecotti, {\sl Int. J. Mod. Phys.}  A6 (1991) 1749
and {\sl Nucl. Phys.}  B355 (1991) 755;
\item{} A. Giveon and D.J. Smit, {\sl Mod. Phys. Lett.}  A6
   24 (1991) 2211.

\item{3.} E. Witten, {\sl Commun. Math. Phys.} 117 (1988) 353;
   118 (1988) 411 and {\sl Nucl. Phys.}  B340
   (1990), 281;
\item{} T. Eguchi and S.K. Yang, {\sl Mod. Phys. Lett.}  A5
   (1900) 1693;
\item{} C. Vafa, {\sl Mod. Phys. Lett.}  A6 (1991) 337;
\item{} K. Li, {\sl Nucl.Phys.} B354 (1991) 711;
\item{} B. Blok and A. Varchenko, {\sl Int.Jou. Mod. Phys.} A7 (1992) 1467;
\item{} R. Dijkgraaf, E. Verlinde and H. Verlinde, {\sl Nucl. Phys.}
    B348 (1991) 435 and  B352 (1991) 59;
\item{} A. Giveon and D.J. Smit, {\sl Progr. Theor. Phys. Suppl.}
102 (1990) 351; {\sl Mod. Phys. Lett.} A6 (1991) 2211; {\sl Int.Jou.Mod.
Phys.} A7 (1992) 973;
\item{} S. Cecotti and C. Vafa, {\sl Nucl. Phys.}  B367 (1991)
   359.

\item{4.} P. Candelas, X.C. de la Ossa, P.S. Green and L. Parkes,
   {\sl Phys. Lett.} 258B (1991)~118;
\item{} P. Candelas, X.C. de la Ossa, P.S. Green and L. Parkes,
{\sl Nucl. Phys.} B359 (1991)~21.

\item{5.} D. Morrison, {\sl Essays on Mirror manifolds},S.T.Yau Editor,
Intenational Press (1992);
\item{} A. Font, {\sl Periods and Duality Symmetries in Calabi--Yau
   Compactifications} (pre\-print UCVFC/DF-1-92);
\item{}  A. Klemm and S. Theisen,
 {\sl Considerations of one modulus Calabi--Yau compactifications:
  Picard--Fuchs equations, K\"ahler potentials and mirror maps}
  (Karlsruhe pre\-print KA-THEP-03-92).

\item{6.} P. Candelas and X.C. de la Ossa,  {\sl Nucl. Phys.}
B355 (1991) 455.

\item{7.} S. Cecotti, S. Ferrara and L. Girardello, {\sl Int.Jou. Mod.
 Phys.} A4 (1989) 2475;
\item{} S. Cecotti, S. Ferrara and L. Girardello,
{\sl Phys. Lett.}  B213 (1988) 443;
\item{} L.J. Dixon, V.S. Kaplunovsky and J. Louis, {\sl Nucl. Phys.}
B329 (1990) 27.

\item{8.} P.Aspinwall and D. Morrison,  {\sl Topological field
   theory and rational curves}, pre\-print {DUK-M-91-12};
\item{} E. Witten,
{\sl Essays on Mirror manifolds }, S.T.Yau Editor,International Press (1992)
\item{} B.R. Greene and M.R. Plesser, {\sl Nucl. Phys.}  B338
   (1990) 15;
\item{} P. Candelas, M. Linker and R. Schimmrigk, {\sl Nucl. Phys.}
    B341 (1990) 383;
\item{} P. Aspinwall, C.A. L\"utken and G.G. Ross, {\sl Phys. Lett.}
 B241 (1990) 373;
\item{} P. Aspinwall, C.A. L\"utken, {\it Nucl. Phys.}  B353
   (1991) 427 and  B355 (1991) 482;
   See also contributions in ``Essay on Mirror Manifolds"
   edited by S.T. Yau, International Press (1992).

\item{9.} B. de Wit and A. Van Proeyen, {\sl Nucl. Phys.} B245 (1984) 89
\item{} B. de Wit, P. Lauwers and A. Van Proeyen, {\sl Nucl. Phys.}
 B255 (1985) 569;
\item{} E. Cremmer, C. Kounnas, A. Van Proeyen,
   J.P. Deredinger, S. Ferrara, B. de Wit and L. Girardello, {\sl Nucl.
   Phys.}  B250 (1985) 385.
\item{10.} A. Strominger, {\sl Commun. Math. Phys.} 133 (1990)  163.

\item{11.} L. Castellani, R. D'Auria and S. Ferrara, {\sl Phys. Lett.} B241
(1990) 57;
\item{} L. Castellani, R. D'Auria and S. Ferrara,
{\sl Class. Quant. Grav.}  1 (1990) 317;
\item{} R. D'Auria, S. Ferrara and P. Fr\`e, {\sl Nucl. Phys.}
 B359 (1991) 705.

\item{12.} K. Kikkawa and M. Yamasaki, {\sl Phys. Lett.} 149B (1984)
357;
\item{} N. Sakai and L. Senda, {\sl Progr. Theor. Phys.}  75
(1986) 692;
\item{} V.P. Nair, A. Shapere, A. Strominger and F. Wilczek, {\sl Nucl.
Phys.}  B287 (1987) 402;
\item{} A. Giveon, E. Rabinovici and G. Veneziano, {\sl Nucl. Phys.}
 B322 (1989) 167;
\item{} A. Shapere and F. Wilczek, {\sl Nucl. Phys.}  B320
(1989) 167;
\item{} M. Dine, P. Huet and N. Seiberg, {\sl Nucl. Phys.}  B322
(1989) 301;
\item{} J. Molera and B. Ovrut, {\sl Phys. Rev.}  D40 (1989)
1150;
\item{} J. Lauer, J. Maas and H.P. Nilles, {\sl Phys. Lett.}
 B226 (1989) 251 and {\sl Nucl. Phys.}  B351 (1991) 353;
\item{} W. Lerche, D. L\"ust and N.P. Warner, {\sl  Phys. Lett.}
 B231 (1989) 418;
\item{} M. Duff, {\sl Nucl. Phys.} B335 (1990) 610;
\item{} A. Giveon and M. Porrati, {\sl Phys. Lett.}  B246
(1990) 54 and {\sl Nucl. Phys.}  B355 (1991) 422;
\item{} Giveon, N. Malkin and E. Rabinovici, {\sl Phys. Lett.}
 B238 (1990) 57;
\item{} J. Erler, D. Jungnickel and H.P. Nilles, {\sl  MPI-Ph/91-90};
\item{} S. Ferrara, D. L\"ust, A. Shapere and S. Theisen, {\sl Phys.
Lett.}  B233 (1989) 147;
\item{} J. Schwarz, {\sl Caltech preprints CALT-6S-1581} (1990)
{\sl CALT-68-1728} (1991) and {\sl CALT-68-1740} (1991);
\item{} J. Erler, D. Jungnickel and H.P. Nilles, {\sl MPI-Ph/91-81}.

\item{13.} E. Witten, {\sl Phys. Lett.} B155 (1985) 151.

\item{14.}  L.J. Dixon, V.S. Kaplunovsky and J. Louis, {\sl Nucl. Phys.}
B355 (1991) 649;
\item{} J.P. Deredinger, S. Ferrara, C. Kounnas and F. Zwirner,
{\sl Nucl. Phys.}  B372 (1992) 145;
\item{} J. Louis, {\sl PASCOS}, (1991), Proceedings, P. Nath Editor, World
Scientific 1991;
\item{} G.L. Cardoso and B. Ovrut, {\sl Nucl. Phys.}  B369
(1992) 351.

\item{15.} A. Cadavid and S. Ferrara, {\sl Phys. Lett.} B267
(1991) 193;
\item{} W. Lerche, D. Smit and N. Warner, {\sl Nucl. Phys.}  B372
(1992) 87;

\item{16.} A. Erd\'elyi, F. Oberhettinger, W. Magnus and F.G. Tricomi,
{\sl Higher Trascendental Functions} Mac Graw--Hill,  New York, 1953.

\item{17.}
P. Fr\`e and P. Soriani, private communication, see also
P. Soriani, SISSA Ph.D--thesis.

\item{18.}  S. Ferrara and J. Louis, {\sl Phys. Lett.} B278
(1992) 240.

\item{19.} A. Ceresole, R. D'Auria, S. Ferrara, W. Lerche and J. Louis,
 {\sl Int.Jou.Mod.Phys.} A8 (1993) 79;

\item{20.} P. Di Francesco, C. Itzynkson and J.B. Zuber, {\sl Commun.
Math. Phys.} 140 (1991) 543.

\item{21.} M. Villasante, {\sl Phys. Rev.} D45 (1992) 1831.

\item{22.}P.Candelas, private communication.

\bye